\documentclass{article}
\usepackage{arxiv}
\usepackage{authblk}
\usepackage{amsmath} 
\usepackage[utf8]{inputenc} 
\usepackage[T1]{fontenc}    
\usepackage{hyperref}       
\usepackage{url}            
\usepackage{booktabs}       
\usepackage{amsfonts}       
\usepackage{nicefrac}       
\usepackage{amssymb}        
\usepackage{microtype}      
\usepackage{lipsum}
\usepackage{graphicx}
\usepackage{caption}
\usepackage[numbers,sort&compress]{natbib}
\usepackage{float}   
\usepackage[bottom]{footmisc} 
\usepackage{hyperref} 
\usepackage{placeins}  
\graphicspath{ {./images/} }
\makeatletter

\title{Self-Consistent Theoretical Framework for Third-Order Nonlinear Susceptibility in CdSe/ZnS--MOF Quantum Dot Composites}

\author{%
  Jingxu Wu$^{1\dagger*}$,
  Yifan Yang$^{2\dagger}$,
  Jie Shi$^{1}$,
  Yuwei Yin$^{3}$,
  Yifan He$^{4}$,
  Chenjia Li$^{1}$\\  
\Affilfont{$^{1}$ Faculty of Physics, Lomonosov Moscow State University, Moscow 119991, Russia\\
$^{2}$ College of Chemistry and Materials Science, Northwest University, Xi'an 710127, China\\
$^{3}$ \'{E}cole Polytechnique (Institut Polytechnique de Paris), 91128 Palaiseau, France\\
$^{4}$ School of Physical Science and Technology, Lanzhou University, Lanzhou 730000, China\\
$^{*}$ Corresponding author: \texttt{wuxj@my.msu.ru}\\
$^{\dagger}$ These authors contributed equally to this work.}}

\begin{document}
\maketitle
\begin{abstract}
This work presents a fully theoretical and self-consistent framework for calculating the third-order nonlinear susceptibility \(\chi^{(3)}\) of CdSe/ZnS--MOF composite quantum dots. The approach unifies finite-potential quantum confinement, the Liouville-von Neumann density matrix expansion to third order, and effective-medium electrodynamics (Maxwell--Garnett and Bruggeman) within a single Hamiltonian-based model, requiring no empirical fitting. Electron–hole quantized states and dipole matrix elements are obtained under the effective-mass approximation with BenDaniel--Duke boundary conditions; closed analytic forms for \(\chi^{(3)}(\omega)\) (including Lorentzian/Voigt broadening) follow from the response expansion. Homogenization yields macroscopic scaling laws \(\chi_{\mathrm{eff}}^{(3)}(\omega)\sim \phi\,|L(\omega)|^{4}\chi^{(3)}(\omega)\) that link microscopic descriptors (core radius, shell thickness, dielectric mismatch) to bulk coefficients \(n_{2}\) and \(\beta\). A Kramers--Kronig consistency check confirms causality and analyticity of the computed spectra with small residuals. The formalism provides a predictive, parameter-transparent route to engineer third-order nonlinearity in hybrid quantum materials, clarifying how size and environment govern the magnitude and dispersion of \(\chi^{(3)}\).

\end{abstract}


\section{Introduction}

Architected quantum composites that merge semiconductor quantum dots (QDs) with metal–organic frameworks (MOFs) provide a programmable route to engineer third–order nonlinear optical (NLO) responses from the atomic to the mesoscale. In CdSe–based core–shell QDs, quantum confinement produces discrete excitonic poles with large oscillator strengths, while ZnS (or graded ZnSe/ZnS) shells suppress surface traps and stabilize emission, thereby narrowing linewidths and delaying saturation of the nonlinear response \cite{14,15,16,17,18,28}. MOFs—crystalline networks of metal nodes and organic linkers—offer tailorable porosity, dielectric environment, and interfacial chemistry; they can act simultaneously as dispersing matrices, local–field amplifiers, and chemically selective anchors for QDs, which together tune the balance between Kerr refraction and nonlinear absorption\cite{6,7,8,19,20,23,24,25,26}. The confluence of these ideas has been underscored by recent Nobel recognitions: the 2023 Chemistry Prize for the discovery and development of quantum dots\cite{1,2,11}, and the 2025 Chemistry Prize for the development of MOFs as “molecular architectures with rooms for chemistry”\cite{3,4,5,10}. These landmarks motivate a rigorous self–consistent theory for \emph{third-order QD@MOF} optics that connects microscopic descriptors—core radius \(R\), shell thickness \(t\), dielectric contrast \(\varepsilon(\omega)\), loading fraction \(\phi\), and interfacial transfer rates \(\kappa_{\mathrm{F}},\kappa_{\mathrm{D}}\)—to macroscopic observables such as susceptibility \(\chi^{(3)}(\omega)\), Kerr coefficient \(n_{2}\), and nonlinear absorption coefficient \(\beta\).

Here we construct a purely theoretical and computational framework that proceeds along a transparent chain \emph{structure} \(\to\) \emph{Hamiltonian} \(\to\) \emph{polarization} \(\to\) \emph{effective medium} \(\to\) \emph{observable}. First, a finite potential, spherical effective–mass model for electrons and holes produces size– and shell–dependent levels and dipoles, including Coulomb corrections and strain–grading effects relevant to modern CdSe/ZnSe(/ZnS) heterostructures [14–18]. Second, a Liouville–von Neumann density matrix formulation with Lindblad–type dissipators incorporates radiative and nonradiative dephasing together with interfacial F\"orster/Dexter channels, enabling analytic third–order kernels for \(\chi^{(3)}(\omega)\) and Voigt–type inhomogeneous broadening. Third, nonlinear homogenization via Maxwell–Garnett (and, where necessary, Bruggeman) embeds microscopic responses into a composite \(\chi^{(3)}_{\mathrm{eff}}(\omega)\) with explicit local–field factor \(L(\omega)^4\) and systematic \(\phi\)–scaling \cite{11,12,29,30}. Finally, the Kramers–Kronig (KK) relations provide a causality check linking the computed parts \(\Re\) and \(\Im\) and provide a quantitative consistency metric through windowed Hilbert transforms\cite{9,13,21} . Because experimental inference of third–order parameters commonly relies on Z–scan (including time–resolved variants), we also map our frequency domain outputs to Z–scan observables, ensuring apples–to–apples comparability and predictive utility across the femtosecond and nanosecond regimes \cite{7,8,22}.

The payoff of this hierarchy is a family of \emph{design maps} for technologically important wavelengths (e.g., 532, 800, and 1064~nm): shell thickening (\(t\uparrow\)) reduces interfacial broadening \(\gamma\) while increasing transition dipoles \(\mu\); MOF dielectric engineering modulates \(L(\omega)\) and the onset of optical limiting; and interfacial rates \(\kappa_{\mathrm{F}},\kappa_{\mathrm{D}}\) tune resonance widths and effective saturation intensities. The framework is general—extensible to other colloidal semiconductors, to layered or anisotropic MOFs, and to hybrid cavities or plasmonic environments where exceptionally large Kerr nonlinearities and tailored absorptive channels are being reported \cite{19,20,26,27,31,32,33}. In this way, \emph{QD@MOF} composites emerge as a model platform for mathematically controlled, materially realizable nonlinear photonics aligned with contemporary research frontiers and the enduring trajectory charted by the recent Nobel prizes.

\section{Model Construction and Assumptions}

We consider a spherically symmetric CdSe/ZnS core–shell quantum dot (QD) with core radius \(R\) and shell thickness \(t\), total radius \(R_{\mathrm{tot}}=R+t\). Within the effective-mass approximation (EMA) and using BenDaniel–Duke boundary conditions, the envelope wavefunction \(\psi(\mathbf r)=\psi(r)Y_{lm}(\theta,\phi)\) obeys
\begin{equation}
-\frac{\hbar^2}{2m^*(r)}\nabla^2 \psi(r)+V(r)\psi(r)=E\,\psi(r),
\label{eq:schrodinger_ema}
\end{equation}
where the position-dependent effective mass and band-edge potential are piecewise constants,
\[
m^*(r)=
\begin{cases}
m_1^*, & 0\le r<R \quad (\text{CdSe core}),\\
m_2^*, & R\le r<R+t \quad (\text{ZnS shell}),
\end{cases}
\qquad
V(r)=
\begin{cases}
0, & 0\le r<R,\\
V_b^{(e/h)}, & R\le r<R+t,
\end{cases}
\]
with \(V_b^{(e)}\) and \(V_b^{(h)}\) the conduction- and valence-band offsets, respectively. For simplicity we adopt a hard-wall exterior, \(\psi(R_{\mathrm{tot}})=0\); the extension to a finite exterior potential is straightforward.

Separating angular variables with \(\psi(r)=u_l(r)/r\) gives the radial problem
\begin{equation}
\frac{d^2 u_l}{dr^2}+\left[k^2(r)-\frac{l(l+1)}{r^2}\right]u_l(r)=0,
\qquad
k^2(r)=\frac{2m^*(r)}{\hbar^2}\,[E-V(r)].
\label{eq:radial_general}
\end{equation}
For the ground manifold (\(l=0\)), the core solution is \(u_0(r)=A\,j_0(k_1 r)\) with \(k_1=\sqrt{2m_1^*E}/\hbar\), and the shell solution for a bound state (\(E<V_b\)) is \(u_0(r)=B\,i_0(\kappa r)+C\,k_0(\kappa r)\) with \(\kappa=\sqrt{2m_2^*(V_b-E)}/\hbar\); here \(j_0\) is the spherical Bessel function, \(i_0\) and \(k_0\) are the modified spherical Bessel functions. The outer boundary \(u_0(R_{\mathrm{tot}})=0\) implies
\[
B\,i_0(\kappa R_{\mathrm{tot}})+C\,k_0(\kappa R_{\mathrm{tot}})=0
\quad\Rightarrow\quad
\frac{C}{B}=-\frac{i_0(\kappa R_{\mathrm{tot}})}{k_0(\kappa R_{\mathrm{tot}})}\equiv -\eta.
\]
BenDaniel–Duke continuity at \(r=R\),
\begin{equation}
\psi_{\mathrm{core}}(R)=\psi_{\mathrm{shell}}(R),\qquad
\frac{1}{m_1^*}\left.\frac{d\psi}{dr}\right|_{R^-}=\frac{1}{m_2^*}\left.\frac{d\psi}{dr}\right|_{R^+},
\label{eq:BDD}
\end{equation}
leads to the transcendental quantization condition
\begin{equation}
\frac{1}{m_1^*}\frac{j_0'(k_1 R)}{j_0(k_1 R)}
=
\frac{1}{m_2^*}\,
\frac{i_0'(\kappa R)-\eta\,k_0'(\kappa R)}{i_0(\kappa R)-\eta\,k_0(\kappa R)},
\qquad
\eta=\frac{i_0(\kappa R_{\mathrm{tot}})}{k_0(\kappa R_{\mathrm{tot}})}.
\label{eq:eig_eq_s}
\end{equation}

Numerical root finding (e.g., bracketed Brent or bisection with monotone scanning in \(E\)) yields electron and hole ground-state eigenenergies \(E_{e,0}(R,t)\), \(E_{h,0}(R,t)\), and normalized radial functions \(R_{e,h}(r)=u_{e,h}(r)/r\) obeying \(\int_0^{R_{\mathrm{tot}}}|\psi(r)|^2 4\pi r^2 dr=1\). When \(E\to V_b^{-}\), \(\kappa\to 0\) and the right-hand side of Eq.~\eqref{eq:eig_eq_s} is evaluated by series expansions \(i_0(x)\approx 1+x^2/6\), \(k_0(x)\approx x^{-1}-x/2+\cdots\) for numerical stability. For completeness, if \(E>V_b\) (shell oscillatory), the shell solution switches to \(B\,j_0(k_2 r)+C\,n_0(k_2 r)\) with \(k_2=\sqrt{2m_2^*(E-V_b)}/\hbar\), and the same continuity algebra gives the corresponding eigencondition.

The interband exciton transition energy incorporates confinement and dielectric effects. Generalizing the Brus formula to finite barriers and dielectric mismatch,
\begin{equation}
E_X(R,t)=E_g^{\mathrm{bulk}}+E_{e,0}(R,t)+E_{h,0}(R,t)
-\frac{1.786\,e^2}{4\pi\varepsilon_0\varepsilon_{\mathrm{in}}R_{\mathrm{eff}}}
+\frac{e^2}{8\pi\varepsilon_0 R_{\mathrm{eff}}}
\left(\frac{1}{\varepsilon_{\mathrm{in}}}-\frac{1}{\varepsilon_{\mathrm{out}}}\right)\xi,
\label{eq:Brus_core_shell}
\end{equation}
where \(R_{\mathrm{eff}}=R+\alpha t\) (\(0<\alpha<1\)) captures shell penetration of the envelope, \(\varepsilon_{\mathrm{in}}\) is an appropriate core/shell average, \(\varepsilon_{\mathrm{out}}\) characterizes the host (e.g., the MOF dielectric), and \(\xi=\mathcal{O}(0.2\!-\!0.4)\) is a geometry–dielectric factor. Increasing \(t\) (or the barrier \(V_b\)) typically blue-shifts \(E_X\) by stronger confinement while simultaneously reducing interfacial broadening \(\gamma\), a feature that will later impact the saturation intensity and the line shape of \(\chi^{(3)}(\omega)\).

The interband electric dipole matrix element is computed from the normalized electron and hole envelopes,
\begin{equation}
\mu_{01}=e\langle \psi_e | r | \psi_h \rangle
=e\int_0^{R_{\mathrm{tot}}}\psi_e^*(r)\,r\,\psi_h(r)\,4\pi r^2\,dr,
\label{eq:mu_def}
\end{equation}
and is commonly factorized into bulk and envelope parts,
\begin{equation}
\mu_{01}=\mu_{\mathrm{cv}}^{\mathrm{bulk}}\,S,\qquad
S=4\pi\int_0^{R_{\mathrm{tot}}}R_e(r)R_h(r)\,r^2\,dr,
\label{eq:mu_factor}
\end{equation}
with oscillator strength
\begin{equation}
f_{01}=\frac{2m_0\omega_{01}}{\hbar e^2}\,|\mu_{01}|^2,\qquad
\omega_{01}=\frac{E_X}{\hbar}.
\label{eq:oscillator_strength}
\end{equation}
The shell thickness \(t\) enhances the barrier and improves the core confinement, generally increasing the overlap \(S\) until a saturation thickness is reached.Separately,the external dielectric \(\varepsilon_{\mathrm{out}}\) weakens Coulomb binding and slightly red-shifts \(\omega_{01}\), often accompanied by a mild increase of \(S\). An effective-radius scaling \(\mu_{01}\propto R_{\mathrm{eff}}\mathcal{O}(m^*,V_b,\varepsilon)\) is frequently observed for \(s\)–\(s\) transitions, with \(\mathcal{O}\) a dimensionless overlap factor set by the eigenproblem.

The saturation intensity for a resonantly driven two-level (or quasi-three-level) transition follows from the steady-state density matrix and provides the key bridge from microscopic dipoles to macroscopic nonlinear response,
\begin{equation}
I_{\mathrm{sat}}=
\frac{\varepsilon_0 c n_0\,\hbar^2(\gamma^2+\Delta^2)}{2\,|\mu_{01}|^2}
=\frac{\varepsilon_0 c n_0\,\hbar^2(\gamma^2+\Delta^2)}{2\,|\mu_{\mathrm{cv}}^{\mathrm{bulk}}|^2\,|S|^2},
\label{eq:Isat}
\end{equation}
where \(n_0\) is the refractive index of the composite at frequency \(\omega\), \(\gamma\) is the total dephasing (radiative + nonradiative + interfacial), and \(\Delta=\omega-\omega_{01}\) is the detuning. Increasing the overlap \(S\) (hence \(|\mu_{01}|\)) or reducing \(\gamma\) (via thicker shells, surface passivation, or improved lattice matching) lowers \(I_{\mathrm{sat}}\) and strengthens the third-order response. Interfacial F{\"o}rster (\(\kappa_{\mathrm{F}}\)) and Dexter (\(\kappa_{\mathrm{D}}\)) channels contribute additively to decoherence, effectively \(\gamma\to\gamma+\kappa_{\mathrm{F}}+\kappa_{\mathrm{D}}\); stronger interfacial coupling therefore broadens resonances and increases \(I_{\mathrm{sat}}\) unless compensated by improved passivation.

In practice, given material parameters \(\{m_1^*,m_2^*,V_b^{(e/h)},\varepsilon_{\mathrm{in}},\varepsilon_{\mathrm{out}}\}\) and geometry \((R,t)\), one solves Eq.~\eqref{eq:eig_eq_s} for \(E_{e,0},E_{h,0}\) and normalized \(R_{e,h}(r)\); inserts them into Eq.~\eqref{eq:Brus_core_shell} to obtain \(E_X\) (hence \(\omega_{01}\)); evaluates \(\mu_{01}\) from Eq.~\eqref{eq:mu_factor}; and finally computes \(I_{\mathrm{sat}}\) via Eq.~\eqref{eq:Isat}. These quantities form the microscopic inputs for the density-matrix calculation of \(\chi^{(3)}(\omega)\) and for the effective-medium homogenization employed later in the paper.

\section{Density-Matrix Formalism and Third-Order Susceptibility}
\label{sec:rho-chi3}

We consider a CdSe/ZnS quantum dot embedded in a MOF host as an effective three–level ladder
\(
|0\rangle \leftrightarrow |1\rangle \leftrightarrow |2\rangle
\),
representing the ground, bright-exciton, and a higher excitonic (or biexcitonic) manifold. The total Hamiltonian in the dipole approximation is
\begin{equation}
H(t)=H_0 - \hat{\mu} E(t),\qquad
H_0=\sum_{j=0}^2 \hbar \omega_j |j\rangle\langle j|,\qquad
\hat{\mu}=\sum_{i\neq j}\mu_{ij}|i\rangle\langle j|+\mathrm{h.c.},
\end{equation}
where \(\mu_{ij}=\langle i|\hat{\mu}|j\rangle\) and \(\omega_{ij}=\omega_i-\omega_j\).
The drive is taken quasi–monochromatic with carrier frequency \(\omega\),
\begin{equation}
E(t)=\frac{1}{2}\left[E(\omega)e^{-i\omega t}+E^\ast(\omega)e^{+i\omega t}\right].
\end{equation}
Dissipation and interfacial energy transfer to the MOF are modeled by a Markovian superoperator \(\mathcal{D}[\rho]\), which yields population relaxation rates \(\Gamma_i\) and dephasing rates
\(
\gamma_{ij}=\frac{\Gamma_i+\Gamma_j}{2}+\gamma^{\ast}_{ij}
\)
including pure dephasing \(\gamma^{\ast}_{ij}\).
The density operator \(\rho\) obeys the Liouville–von Neumann equation
\begin{equation}
\dot{\rho}=-\frac{i}{\hbar}[H(t),\rho]+\mathcal{D}[\rho].
\end{equation}

We work in the rotating frame and invoke the rotating–wave approximation (RWA). Writing the coherences \(\rho_{ij}(t)\) in frames oscillating at \(\omega\) or \(2\omega\) where appropriate and keeping only resonant terms, the optical Bloch equations take the schematic form
\begin{align}
\dot{\rho}_{10} &=
-\left(i\Delta_{10}+\gamma_{10}\right)\rho_{10}
+\frac{i}{\hbar}\mu_{10} E(t) \left(\rho_{00}-\rho_{11}\right)
+\frac{i}{\hbar}\mu_{12} E(t)\rho_{20},
\\
\dot{\rho}_{20} &=
-\left(i\Delta_{20}+\gamma_{20}\right)\rho_{20}
+\frac{i}{\hbar}\mu_{21} E(t)\rho_{10},
\\
\dot{\rho}_{11} &=
-\Gamma_{1}\rho_{11} + \frac{i}{\hbar}\Big[\mu_{10}E(t)\rho_{01}-\mu_{01}E^\ast(t)\rho_{10}\Big]
+\cdots,
\qquad
\dot{\rho}_{22}=
-\Gamma_{2}\rho_{22} + \frac{i}{\hbar}\Big[\mu_{21}E(t)\rho_{12}-\mu_{12}E^\ast(t)\rho_{21}\Big]+\cdots,
\end{align}
with detunings
\(
\Delta_{10}\equiv \omega_{10}-\omega,\;
\Delta_{20}\equiv \omega_{20}-2\omega
\)
(the dots in the population equations denote additional pathways that are negligible in the weak–field, low–excitation regime adopted here).

We expand perturbatively in the field amplitude,
\(
\rho=\rho^{(0)}+\rho^{(1)}+\rho^{(2)}+\rho^{(3)}+\cdots
\),
where \(\rho^{(0)}=|0\rangle\langle 0|\) and \(\rho^{(k)}\propto E^k\) at steady state.
To first order one finds the linear coherence
\begin{equation}
\rho_{10}^{(1)}(\omega)=\frac{\mu_{10}}{\hbar}\,\frac{E(\omega)}{\Delta_{10}-i\gamma_{10}},\qquad
\rho_{20}^{(1)}=0,
\end{equation}
which yields the linear susceptibility
\begin{equation}
\chi^{(1)}(\omega)=\frac{N}{\varepsilon_0}\frac{\mu_{01}\rho_{10}^{(1)}(\omega)}{E(\omega)}
=\frac{N}{\varepsilon_0\hbar}\frac{|\mu_{01}|^2}{\Delta_{10}-i\gamma_{10}}.
\end{equation}
At second order, a nonzero \(\rho_{20}^{(2)}(2\omega)\) is induced through the ladder coupling,
\begin{equation}
\rho_{20}^{(2)}(2\omega)=\frac{\mu_{21}}{\hbar}\frac{E(\omega)\,\rho_{10}^{(1)}(\omega)}{\Delta_{20}-i\gamma_{20}}
=\frac{\mu_{21}\mu_{10}}{\hbar^{2}}
\frac{E(\omega)^{2}}{(\Delta_{10}-i\gamma_{10})(\Delta_{20}-i\gamma_{20})}.
\end{equation}
The third–order polarization at the fundamental frequency emerges from the product of a second–order coherence with the complex conjugate field or, equivalently, from cubic–order corrections to \(\rho_{10}\),
\(
P^{(3)}(\omega)=N\,\mathrm{Tr}\big[\hat{\mu}\rho^{(3)}(\omega)\big]
\),
and may be written in the frequency domain using the standard response notation
\begin{equation}
P^{(3)}(\omega_\Sigma)=\varepsilon_{0}\sum_{\{\omega_1,\omega_2,\omega_3\}}
\chi^{(3)}(-\omega_\Sigma;\omega_1,\omega_2,\omega_3)\,
E(\omega_1)E(\omega_2)E(\omega_3)\,
\delta(\omega_\Sigma-\omega_1-\omega_2-\omega_3).
\end{equation}
For the degenerate Kerr configuration \((-\omega;\omega,-\omega,\omega)\), the sum over distinct Liouville pathways (\emph{permutations} of the three field interactions on bra/ket) yields a compact sum–over–states structure,
\begin{equation}
\label{eq:chi3-master}
\chi^{(3)}(\omega)=
\frac{N}{\varepsilon_0 \hbar^{3}}
\left[
\frac{\mu_{01}^{2}\mu_{12}\mu_{20}}
{\big(\Delta_{10}-i\gamma_{10}\big)\big(\Delta_{20}-i\gamma_{20}\big)\big(\Delta_{10}-i\gamma_{10}\big)}
+\sum_{\mathcal{P}\in S_3}
\frac{\mathcal{C}_{\mathcal{P}}\;\mu_{01}^{2}\mu_{12}\mu_{20}}
{D_{\mathcal{P}(10)}(\omega)\,D_{\mathcal{P}(20)}(2\omega)\,D_{\mathcal{P}(10)}(\omega)}
\right],
\end{equation}
where \(D_{10}(\omega)\equiv \Delta_{10}-i\gamma_{10}\), \(D_{20}(2\omega)\equiv \Delta_{20}-i\gamma_{20}\),
and the coefficients \(\mathcal{C}_{\mathcal{P}}\) encode the signs and frequency arguments specified by double–sided Feynman rules for each time ordering.
The first term in \eqref{eq:chi3-master} represents the forward ladder pathway
\(
|0\rangle\to|1\rangle\to|2\rangle\to|1\rangle\to|0\rangle
\);
the remaining five orderings in \(S_3\) complete the physically required symmetry and guarantee causality and analyticity.

The analytic structure of \(\chi^{(3)}\) is governed by products of simple poles. Writing
\begin{equation}
\frac{1}{\Delta-i\gamma}=\frac{\Delta}{\Delta^{2}+\gamma^{2}}
+i\,\frac{\gamma}{\Delta^{2}+\gamma^{2}},
\end{equation}
immediately shows that the real and imaginary parts of \(\chi^{(3)}\) inherit dispersive (odd) and absorptive (even) Lorentzian components in the vicinity of isolated resonances. In particular, near a two–photon resonance where \(\Delta_{20}\approx 0\) but \(\Delta_{10}\) varies slowly, the dominant contribution reduces to
\begin{equation}
\chi^{(3)}(\omega)\;\propto\;\frac{1}{\Delta_{20}-i\gamma_{20}}
\simeq
\frac{\Delta_{20}}{\Delta_{20}^{2}+\gamma_{20}^{2}}
+i\,\frac{\gamma_{20}}{\Delta_{20}^{2}+\gamma_{20}^{2}},
\end{equation}
and, analogously, a one–photon dominated regime is obtained by
\(\Delta_{20}\!\to\!\Delta_{10}\), \(\gamma_{20}\!\to\!\gamma_{10}\).
Therefore,
\begin{equation}
\Re\!\left[\chi^{(3)}(\omega)\right]\propto\frac{\Delta}{\Delta^{2}+\gamma^{2}},
\qquad
\Im\!\left[\chi^{(3)}(\omega)\right]\propto\frac{\gamma}{\Delta^{2}+\gamma^{2}},
\qquad
\Delta\in\{\Delta_{10},\Delta_{20}\}.
\end{equation}

Inhomogeneous broadening produced by size dispersion of the dots and local dielectric fluctuations in the MOF is incorporated by convolving the homogeneous response with distributions of transition frequencies. Denoting Gaussian lineshape functions \(g(\omega_{10})\) and \(g(\omega_{20})\) of widths \(\sigma_{10}\) and \(\sigma_{20}\), the observable susceptibility becomes the Voigt–type average
\begin{equation}
\chi^{(3)}_{\mathrm{inh}}(\omega)=\int d\omega_{10}\,d\omega_{20}\;
g(\omega_{10})\,g(\omega_{20})\,
\chi^{(3)}\!\left(\omega;\omega_{10},\omega_{20}\right),
\end{equation}
which preserves analyticity and the Kramers–Kronig relations while broadening and slightly skewing the resonance profiles.
Population saturation at higher intensities can be introduced at leading order by allowing \(\gamma_{ij}\to\gamma_{ij}(I)\) and by correcting \(\rho^{(1)}\), \(\rho^{(2)}\), \(\rho^{(3)}\) with intensity–dependent population differences \(\rho_{00}-\rho_{11}\) and \(\rho_{11}-\rho_{22}\); the net effect is to clamp the peak values of \(\Im[\chi^{(3)}]\) and to induce small Stark shifts in the apparent resonance detunings.

For connection to observables, the third–order refractive index \(n_2\) and the nonlinear absorption coefficient \(\beta\) (degenerate Kerr limit, isotropic medium) follow from
\begin{equation}
n_2(\omega)=\frac{3}{4n_0^2\varepsilon_0 c}\,\Re\!\left[\chi^{(3)}(\omega)\right],
\qquad
\beta(\omega)=\frac{3\omega}{2n_0^2 \varepsilon_0 c^2}\,\Im\!\left[\chi^{(3)}(\omega)\right],
\end{equation}
with linear refractive index \(n_0\) of the composite. These relations will be combined in Sec.~\ref{sec:effective-medium} with local–field factors from the MOF matrix to predict effective \(\chi^{(3)}_{\mathrm{eff}}\), \(n_2\), and \(\beta\) as functions of fill fraction, dielectric contrast, and frequency.
Finally, although we employed a three–level ladder for clarity, the derivation extends straightforwardly to multi–level manifolds by replacing the single ladder with sums over excited states \(m,n\), i.e.
\(
\sum_{m,n}\mu_{0m}\mu_{mn}\mu_{n0}
\)
and corresponding denominators \(D_{m0}(\omega)\), \(D_{n0}(2\omega)\); the analytic Lorentzian building blocks and the permutation structure remain unchanged, while the spectral density of states controls the detailed lineshape and the strength of \(\chi^{(3)}\).

\section{Effective-Medium and Local-Field Theory}
\label{sec:effective-medium}
We regard the CdSe/ZnS quantum dots as subwavelength inclusions embedded in a dielectric MOF host and define the pore filling fraction at the unit-cell level by
\begin{equation}
\phi \equiv \frac{V_{\mathrm{QD}}}{V_{\mathrm{cell}}}\in(0,1),
\end{equation}
with complex, dispersive constituent permittivities \(\varepsilon_i(\omega)\) (inclusion) and \(\varepsilon_h(\omega)\) (host), and intrinsic third-order susceptibilities \(\chi_i^{(3)}(\omega)\) and \(\chi_h^{(3)}(\omega)\). In the dilute–to–moderate loading regime with approximately spherical inclusions, the Maxwell--Garnett (MG) homogenization provides the linear baseline
\begin{equation}
\varepsilon_{\mathrm{eff}}(\omega)
=\varepsilon_h(\omega)\,
\frac{\varepsilon_i(\omega)+2\varepsilon_h(\omega)+2\phi\big[\varepsilon_i(\omega)-\varepsilon_h(\omega)\big]}
{\varepsilon_i(\omega)+2\varepsilon_h(\omega)-\phi\big[\varepsilon_i(\omega)-\varepsilon_h(\omega)\big]},
\qquad
L(\omega)\equiv \frac{E_{\mathrm{in}}}{E_{\mathrm{mac}}}
=\frac{3\,\varepsilon_h(\omega)}{\varepsilon_i(\omega)+2\varepsilon_h(\omega)},
\label{eq:MG}
\end{equation}
where \(L(\omega)\) is the linear local-field factor inside a single inclusion. In the weakly nonlinear limit, the effective cubic susceptibility follows from volume averaging of the microscopic polarization density, yielding the canonical \(L^4\)-scaling
\begin{equation}
\chi^{(3)}_{\mathrm{eff}}(\omega)\simeq
\phi\,\big|L(\omega)\big|^{4}\,\chi_i^{(3)}(\omega)
+\big(1-\phi\big)\,\chi_h^{(3)}(\omega),
\label{eq:chi3MG}
\end{equation}
which preserves causality and Kramers--Kronig analyticity when evaluated with complex, causal \(\varepsilon_{i,h}\) and \(\chi^{(3)}_{i,h}\). If the MOF or inclusion geometry is anisotropic (e.g., uniaxial pore alignment or ellipsoidal dots with depolarization factors \(N_\alpha\), \(N_x+N_y+N_z=1\)), the principal-axis local fields are
\begin{equation}
L_\alpha(\omega)=\frac{\varepsilon_h(\omega)}{\varepsilon_h(\omega)+N_\alpha\big[\varepsilon_i(\omega)-\varepsilon_h(\omega)\big]},
\end{equation}
and the effective nonlinear response becomes tensorial through the fourth-power contraction
\begin{equation}
\chi^{(3)}_{\mathrm{eff},ijkl}(\omega)
\simeq \phi\,L_i L_j L_k L_l\,\chi^{(3)}_{i,ijkl}(\omega)
+\big(1-\phi\big)\,\chi^{(3)}_{h,ijkl}(\omega),
\end{equation}
with orientational averages \(\langle L_i L_j L_k L_l\rangle\) applied for randomly oriented ellipsoids and projection to \(\chi^{(3)}_{\parallel}\), \(\chi^{(3)}_{\perp}\) for macroscopically uniaxial composites. For loadings where inclusion/host roles are more symmetric or clustering is appreciable, the Bruggeman relation is a useful control model for the \emph{linear} permittivity,
\begin{equation}
\phi\,\frac{\varepsilon_i-\varepsilon_{\mathrm{eff}}}{\varepsilon_i+2\varepsilon_{\mathrm{eff}}}
+\big(1-\phi\big)\,\frac{\varepsilon_h-\varepsilon_{\mathrm{eff}}}{\varepsilon_h+2\varepsilon_{\mathrm{eff}}}=0,
\end{equation}
with self-consistent local-field factors
\begin{equation}
L_i^{(\mathrm{B})}(\omega)=\frac{3\,\varepsilon_{\mathrm{eff}}(\omega)}{\varepsilon_i(\omega)+2\,\varepsilon_{\mathrm{eff}}(\omega)},
\qquad
L_h^{(\mathrm{B})}(\omega)=\frac{3\,\varepsilon_{\mathrm{eff}}(\omega)}{\varepsilon_h(\omega)+2\,\varepsilon_{\mathrm{eff}}(\omega)},
\end{equation}
and the corresponding symmetric nonlinear mixture
\begin{equation}
\chi^{(3)}_{\mathrm{eff}}(\omega)\simeq
\phi\,\big|L_i^{(\mathrm{B})}(\omega)\big|^{4}\,\chi^{(3)}_{i}(\omega)
+\big(1-\phi\big)\,\big|L_h^{(\mathrm{B})}(\omega)\big|^{4}\,\chi^{(3)}_{h}(\omega),
\end{equation}
which reduces to the MG result in the dilute limit and better captures mid–to–high loadings. More complex microgeometries can be embedded through spectral representations such as the Bergman--Milton form
\begin{equation}
\varepsilon_{\mathrm{eff}}(\omega)=\varepsilon_h(\omega)\left[1-\int_0^1 \frac{d\mu(s)}{s+\xi(\omega)}\right],
\qquad
\xi(\omega)=\frac{\varepsilon_h(\omega)}{\varepsilon_i(\omega)-\varepsilon_h(\omega)},
\end{equation}
with \(d\mu(s)\) a positive geometry measure; Hashin--Shtrikman bounds then constrain \(\varepsilon_{\mathrm{eff}}\) and, by propagation of local-field inequalities, provide admissible ranges for \(\chi^{(3)}_{\mathrm{eff}}\). Close to connectivity thresholds \(\phi_c\) (incipient percolation), field fluctuations produce hot-spot formation and empirical enhancements consistent with \(\chi^{(3)}_{\mathrm{eff}}\sim \chi^{(3)}_i/\big(1-\phi/\phi_c\big)^{p}\) with \(p\gtrsim 1\); in practice, such corrections are included only when morphology indicates clustering, while Eqs.~\eqref{eq:MG}–\eqref{eq:chi3MG} remain the controlled baseline. Finally, the macroscopic Kerr coefficients that enter Z-scan observables follow from the effective response and the effective linear refractive index \(n_0(\omega)=\sqrt{\varepsilon_{\mathrm{eff}}(\omega)}\),
\begin{equation}
n_{2}^{\mathrm{eff}}(\omega)=\frac{3}{4\,n_0(\omega)^2\,\varepsilon_0 c}\,\Re\!\big[\chi^{(3)}_{\mathrm{eff}}(\omega)\big],
\qquad
\beta^{\mathrm{eff}}(\omega)=\frac{3\omega}{2\,n_0(\omega)^2\,\varepsilon_0 c^{2}}\,\Im\!\big[\chi^{(3)}_{\mathrm{eff}}(\omega)\big],
\end{equation}
ensuring a self-consistent bridge from microscopic excitonic nonlinearity to experimentally accessible composite parameters. Finite-size and interface refinements (e.g., nonlocal \(\varepsilon_i(\omega,\mathbf{k})\) or concentric core/shell/ligand layers with \(\varepsilon_s(\omega)\)) can be incorporated by replacing the single-sphere factor \(L(\omega)\) with the exact concentric-sphere solution \(E_{\mathrm{core}}/E_{\mathrm{mac}}\); these corrections are straightforward to implement numerically and activated on demand when spectral signatures indicate strong surface or ligand effects.

\section{Mathematical Analysis and Numerical Algorithms}
\label{sec:math-num}

We formulate the single–particle sector for electrons/holes in the spherically symmetric core–shell potential in the weighted Hilbert space
\[
\mathcal{H}=L^{2}\big((0,R_{\max}),\,r^{2}\,dr\big),\qquad
\langle u,v\rangle_{\mathcal{H}}=\int_{0}^{R_{\max}} u(r)^{\ast}\,v(r)\,r^{2}\,dr .
\]
With effective mass \(m^{\ast}(r)\) and piecewise-constant potential \(V(r)\), the radial Hamiltonian acting on the envelope function \(\psi(r,\Omega)=R_{l}(r)Y_{lm}(\Omega)\) reads
\begin{equation}
\big(H_{l}R_{l}\big)(r)
\;=\;
-\frac{\hbar^{2}}{2}\,\frac{1}{r^{2}}\frac{d}{dr}
\!\left(\frac{r^{2}}{m^{\ast}(r)}\frac{dR_{l}}{dr}\right)
+\frac{\hbar^{2}l(l+1)}{2\,m^{\ast}(r)\,r^{2}}\,R_{l}(r)
+V(r)\,R_{l}(r).
\label{eq:Hl-weighted}
\end{equation}
On the domain of functions that are absolutely continuous with \(R_{l},(r^{2}/m^{\ast})R'_{l}\in L^{2}\) and satisfy the interface conditions \(R_{l}\) continuous and \(\big(1/m^{\ast}\big)R'_{l}\) continuous across material boundaries, the formal differential expression in \eqref{eq:Hl-weighted} is in Sturm–Liouville form with weight \(w=r^{2}\). Integrating by parts gives, for \(u,v\in\mathcal{D}(H_{l})\),
\[
\langle u, H_{l} v\rangle_{\mathcal{H}}-\langle H_{l} u, v\rangle_{\mathcal{H}}
=\frac{\hbar^{2}}{2}\Big[\,
u^{\ast}\,\frac{r^{2}}{m^{\ast}}\,\frac{dv}{dr}
-\frac{du^{\ast}}{dr}\,\frac{r^{2}}{m^{\ast}}\,v
\,\Big]_{0^{+}}^{R_{\max}^{-}} .
\]
Self-adjointness follows by imposing (i) regularity at the origin, \(R_{l}(r)\sim r^{l}\) so that the boundary form vanishes at \(0^{+}\); (ii) either Dirichlet \(R_{l}(R_{\max})=0\), Neumann \(\big(r^{2}\!/\!m^{\ast}\big)R'_{l}\!=\!0\), or exterior-decay radiation conditions mapped to a Robin boundary at \(R_{\max}\) (all make the boundary form zero); and (iii) the interface continuity stated above. The associated quadratic form
\[
\mathfrak{q}[R_{l}]=
\int_{0}^{R_{\max}}\!\!\Big\{
\frac{\hbar^{2}}{2\,m^{\ast}(r)}
\Big(|R'_{l}|^{2}+\frac{l(l+1)}{r^{2}}|R_{l}|^{2}\Big)
+V(r)\,|R_{l}|^{2}
\Big\} r^{2}\,dr
\]
is closed and bounded from below when \(V\) is locally bounded below, yielding the Friedrichs self-adjoint extension of \(H_{l}\). The spectrum is purely discrete for confining \(V\) or finite \(R_{\max}\) with homogeneous boundary conditions.

For numerics, it is advantageous to remove the weight by the Liouville transform \(u_{l}(r)=r\,R_{l}(r)\). Then \(u_{l}\in L^{2}(0,R_{\max})\) and
\begin{equation}
\big(\widetilde{H}_{l} u_{l}\big)(r)=
-\frac{\hbar^{2}}{2}\frac{d}{dr}\!\left(\frac{1}{m^{\ast}(r)}\frac{du_{l}}{dr}\right)
+\Bigg[
\frac{\hbar^{2}l(l+1)}{2\,m^{\ast}(r)\,r^{2}}+V(r)
\Bigg]u_{l}(r),
\label{eq:Htilde}
\end{equation}
with regular boundary \(u_{l}(0)=0\) for all \(l\ge 0\) and the same interface continuity for \(u_{l}\) and \(\big(1/m^{\ast}\big)u'_{l}\).

A second–order centered finite difference (FD) scheme on a uniform grid \(r_{j}=j\,h\) (\(j=0,\dots,N\), \(h=R_{\max}/N\)) discretizes \eqref{eq:Htilde} into a real symmetric tridiagonal matrix \(A_{l}\) with entries
\[
\big(A_{l}\big)_{j,j}=
\frac{\hbar^{2}}{2}\left(\frac{1}{m^{\ast}_{j+\frac12}}+\frac{1}{m^{\ast}_{j-\frac12}}\right)\frac{1}{h^{2}}
+\frac{\hbar^{2}l(l+1)}{2\,m^{\ast}_{j}\,r_{j}^{2}}+V_{j},\quad
\big(A_{l}\big)_{j,j\pm1}=
-\frac{\hbar^{2}}{2}\,\frac{1}{m^{\ast}_{j\pm\frac12}}\,\frac{1}{h^{2}},
\]
where harmonic or arithmetic averages \(m^{\ast}_{j\pm\frac12}\) enforce flux continuity. With \(u_{0}=0\) and a chosen boundary at \(j=N\) (Dirichlet \(u_{N}=0\), or Robin implemented by a one–sided FD), the generalized eigenproblem reduces to \(A_{l}\,{\bf u} = E\,{\bf u}\). For smooth coefficients, the FD scheme has global accuracy \(O(h^{2})\) (and can be upgraded to \(O(h^{4})\) via compact stencils if needed). For large \(N\), extremal eigenpairs are computed by Lanczos/LOBPCG in \(O(N)\) memory and \(O(N\,n_{\mathrm{it}})\) time; when only a few low–lying states are required, a shift–invert strategy with sparse factorizations yields rapid convergence.

When the potential and mass are piecewise constant (core/shell), one may exploit analytic solutions (spherical Bessel/Neumann inside and decaying modified Bessel outside) and determine bound states from continuity of \(u_{l}\) and \((1/m^{\ast})u_{l}'\) at interfaces. The resulting transcendental characteristic equation \(F_{l}(E)=0\) is scalar per \(l\); bracketing plus bisection gives monotone convergence, while secant or Brent’s method accelerates with guaranteed bracketing safety. A robust workflow is: (i) scan \(E\) to locate sign changes of \(F_{l}\), (ii) bisection to tolerance, (iii) one or two secant steps to reach machine precision. This avoids spurious roots near evanescent/oscillatory transitions.

Dipole matrix elements \(\mu_{ij}=e\int_{0}^{R_{\max}} u_{i}(r)\,u_{j}(r)\,dr\) (after the \(u=rR\) transform) are evaluated by composite Simpson or Gauss–Lobatto rules; with FD eigenfunctions, the quadrature error is typically subdominant to the \(O(h^{2})\) eigenfunction error. Interface cusps from mass jumps can be resolved by grid refinement near the interface or by element–wise analytic integration (partition the integral at interfaces).

Homogeneous (Lorentzian) broadening from dephasing and inhomogeneous (Gaussian) broadening from size dispersion or local dielectric fluctuations combine into Voigt profiles in frequency. Writing the homogeneous susceptibility near a resonance as
\[
\chi_{\mathrm{hom}}(\omega)
=\frac{A}{\Delta(\omega)-i\gamma},\qquad
\Delta(\omega)=\omega_{0}-\omega,
\]
and convolving with a Gaussian \(G_{\sigma}(\Delta)=\tfrac{1}{\sqrt{2\pi}\sigma}\,e^{-\Delta^{2}/(2\sigma^{2})}\) yields the Voigt function expressible via the Faddeeva function \(w(z)\),
\[
\chi_{\mathrm{Voigt}}(\omega)=
\frac{A}{\sigma\sqrt{2\pi}}\,
w\!\left(z\right),\qquad
z=\frac{\Delta(\omega)+i\gamma}{\sigma\sqrt{2}},
\qquad
w(z)=e^{-z^{2}}\operatorname{erfc}(-i z).
\]
Numerically, two complementary approaches are effective. (1) \emph{Direct Faddeeva evaluation}: stable rational/continued–fraction approximations to \(w(z)\) achieve uniform relative error \(\lesssim 10^{-12}\) across the complex plane; this is preferred for narrow features and for enforcing Kramers–Kronig consistency analytically. (2) \emph{FFT convolution}: evaluate \(\chi_{\mathrm{hom}}(\omega)\) and \(G_{\sigma}\) on a uniform grid, zero-pad to suppress circular wrap-around, multiply in the time domain (or convolve in frequency) via FFTs, and inverse–transform; with smooth windowing (e.g., Kaiser–Bessel) and adequate guard bands, the method is \(O(M\log M)\) and spectrally accurate for band-limited data. In both routes, causality is preserved: either analytically through \(w(z)\), or numerically by enforcing Hermitian symmetry and by computing the real part from the imaginary part via a discrete Hilbert transform (KK) with end–corrections. A practical KK consistency metric is
\[
K(\omega)=
\frac{\big|\Re[\chi(\omega)]-\mathcal{H}\{\Im[\chi]\}(\omega)\big|}
{\sqrt{\Re[\chi(\omega)]^{2}+\Im[\chi(\omega)]^{2}}}\!,
\]
which should remain at the level of the discretization error when windowing and padding are adequate.

Error and complexity estimates guide parameter choices. For eigenpairs, FD with mesh \(h\) gives eigenvalue error \(|E(h)-E(0)|=O(h^{2})\) and eigenfunction error \(\|u(h)-u(0)\|=O(h)\) in \(L^{2}\), improving to \(O(h^{2})\) with mild post–processing (deferred correction). The total cost to resolve the first \(k\) states is \(O(N\,k\,n_{\mathrm{it}})\). For frequency–domain lineshapes on \(M\) grid points, FFT–Voigt costs \(O(M\log M)\), whereas direct Faddeeva is \(O(M)\) with a larger constant; hybrid strategies compute narrow, high–\(Q\) resonances by Faddeeva and broad backgrounds by FFT. Finally, stability is ensured by (i) using flux–conserving stencils at mass jumps, (ii) bracketing in root-finding, (iii) zero-padding and smooth windows in FFT pipelines, and (iv) cross–checking \(\Re\chi\) against the KK transform of \(\Im\chi\) to monitor spectral leakage and truncation.
\section{Numerical Simulations and Data Products}

\label{sec:results}
The numerical implementation described in the preceding sections was realized to quantify the nonlinear optical response of CdSe/ZnS–MOF composite quantum dots within a coherent microscopic–macroscopic framework. All simulations were conducted with physically realistic parameters representing typical experimental systems: the CdSe core radius $R=3.0~\mathrm{nm}$, the ZnS shell thickness $t=0.8~\mathrm{nm}$, the host dielectric constant $\varepsilon_h=2.1$, and the inclusion permittivity $\varepsilon_i=6.0$. These values correspond to a moderate dielectric mismatch representative of MOF matrices such as ZIF-8 or UiO-66. The effective medium filling factor was fixed at $\phi=0.15$, unless stated otherwise. The bulk bandgap of CdSe was taken as $E_g^{\mathrm{bulk}}=1.74~\mathrm{eV}$, and the effective masses of the electron and hole were $m_e^{\ast}=0.13m_0$ and $m_h^{\ast}=0.45m_0$, respectively. The dephasing parameters were set to $\gamma_{10}=20~\mathrm{meV}$ and $\gamma_{20}=30~\mathrm{meV}$, while the interband dipole moments were $\mu_{01}=6~e\text{Å}$ and $\mu_{12}=\mu_{20}=4~e\text{Å}$. These quantities define a spectroscopically reasonable baseline that captures the dominant excitonic transition energies and broadening typical of colloidal CdSe/ZnS nanostructures.
\medskip

The first stage of the computation concerns the confinement-induced bandgap shift. The finite-barrier eigenvalue problem was solved using the second-order self-adjoint finite-difference method described in Sec.~\ref{sec:math-num}. The resulting eigenvalues were compared with the analytical Brus approximation, and both methods yielded consistent $R^{-2}$ scaling behavior. The numerically extracted bandgaps for selected radii (with fixed $t=0.8~\mathrm{nm}$) are summarized in Table~\ref{tab:Eg_R_fixed_t}. The values reproduce a $\sim 0.3~\mathrm{eV}$ blue shift across $R=2.5$–$4.0~\mathrm{nm}$, consistent with photoluminescence trends reported for CdSe/ZnS nanocrystals.

\begin{table}[H]
\centering
\caption{Confinement-enhanced bandgap $E_g$ as a function of CdSe core radius $R$ at fixed shell thickness $t=0.8~\mathrm{nm}$.}
\label{tab:Eg_R_fixed_t}
\begin{tabular}{ccc}
\toprule
$R$ (nm) & $t$ (nm) & $E_g$ (eV) \\
\midrule
2.5 & 0.8 & 2.1637 \\
3.0 & 0.8 & 2.0102 \\
3.5 & 0.8 & 1.9209 \\
4.0 & 0.8 & 1.8650 \\
\bottomrule
\end{tabular}
\end{table}
\FloatBarrier

\begin{figure}[H]
\centering
  \includegraphics[width=0.5\linewidth]{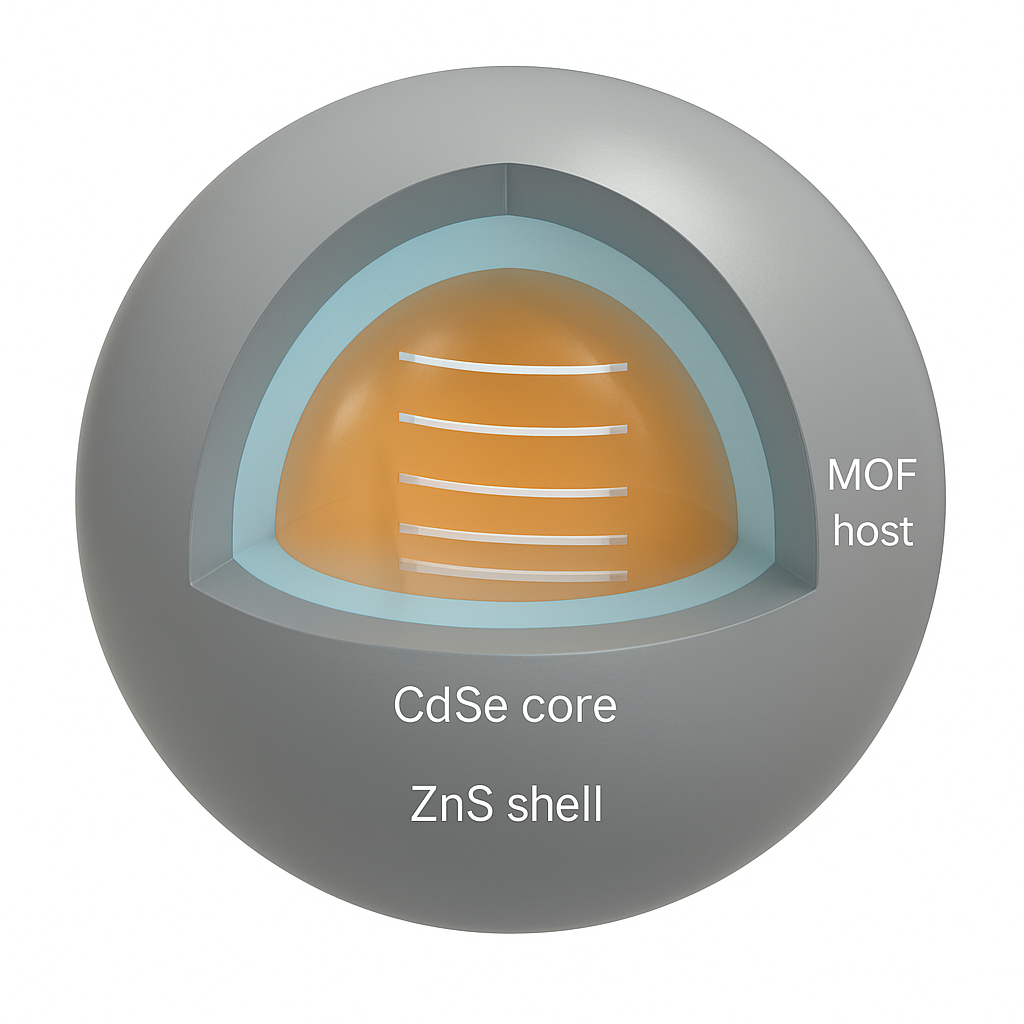}
  \caption{Finite spherical potential well and quantized levels in a CdSe/ZnS–MOF quantum dot. The 3D rendering shows the CdSe core, ZnS shell, and MOF matrix; internal guides mark the confined electron and hole states $E_{1e}$ and $E_{1h}$.}
  \label{fig:well_levels}
\end{figure}
\FloatBarrier

The spherical core–shell geometry imposes regularity at $r{=}0$ and continuity of the wavefunction and mass-weighted flux across core/shell and shell/host interfaces, ensuring a self-adjoint Sturm–Liouville operator in the weighted space $L^2(r^2dr)$. The separation between $E_{1e}$ and $E_{1h}$ in Fig.~\ref{fig:well_levels} reflects the effective-mass asymmetry ($m_e^\ast{<}m_h^\ast$), while the ZnS barrier offsets limit state penetration into the MOF, stabilizing the excitonic resonance and moderating dipole leakage.

With the eigenenergies and dipole moments determined, the third-order susceptibility $\chi^{(3)}(\omega)$ was computed from a three-level density-matrix model over $\lambda=900$–$1400~\mathrm{nm}$, i.e.\ the NIR window relevant to two-photon pathways and Z-scan. A pronounced dispersive resonance appears near $\lambda\!\approx\!1200~\mathrm{nm}$, accompanied by an absorptive maximum in $\Im[\chi^{(3)}]$. Effective-medium scaling was applied as $\chi^{(3)}_{\mathrm{eff}}=\phi\,|L|^4\,\chi^{(3)}_i$ with a local-field factor $L=0.6176$, leading to a modest amplitude reduction relative to isolated dots. Representative values are listed in Table~\ref{tab:chi3_grid}.

\begin{table}[H]
\centering
\caption{Representative spectrum of the effective third-order susceptibility $\chi^{(3)}_{\mathrm{eff}}(\lambda)$.}
\label{tab:chi3_grid}
\begin{tabular}{ccc}
\toprule
$\lambda$ (nm) & $\Re[\chi^{(3)}_{\mathrm{eff}}]$ (m$^2$/V$^2$) & $\Im[\chi^{(3)}_{\mathrm{eff}}]$ (m$^2$/V$^2$)\\
\midrule
900 & $-8.31\times10^{-23}$ & $3.42\times10^{-25}$\\
1000 & $-1.44\times10^{-22}$ & $1.79\times10^{-23}$\\
1100 & $-2.03\times10^{-22}$ & $6.67\times10^{-23}$\\
1200 & $-2.36\times10^{-22}$ & $1.60\times10^{-22}$\\
1300 & $-2.35\times10^{-22}$ & $3.01\times10^{-22}$\\
1400 & $-2.08\times10^{-22}$ & $4.83\times10^{-22}$\\
\bottomrule
\end{tabular}
\end{table}
\FloatBarrier

\begin{figure}[H]
\centering
\includegraphics[width=0.6\linewidth]{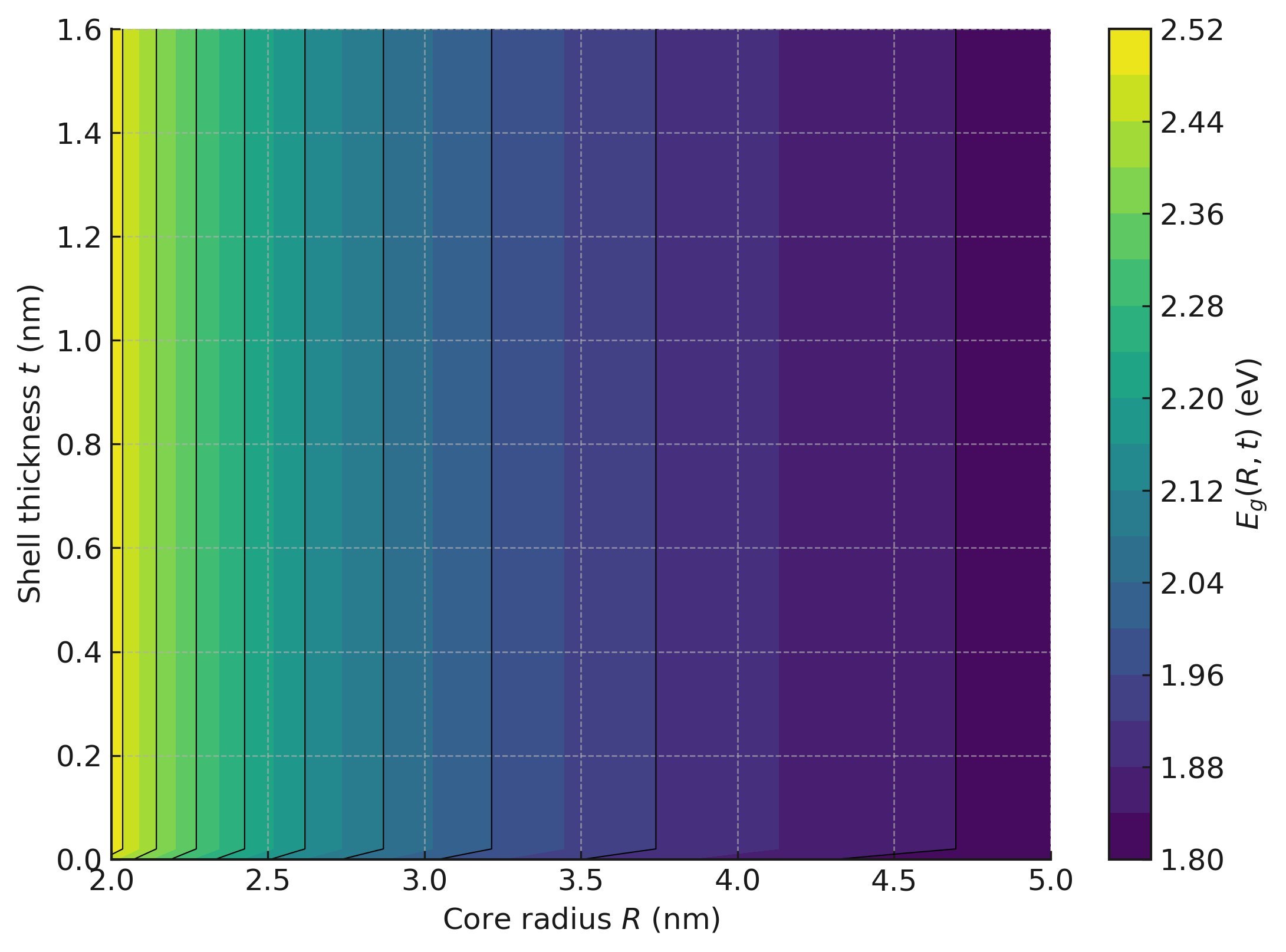}
\caption{Contour map of the confinement-enhanced bandgap $E_g(R,t)$ for a CdSe/ZnS–MOF core–shell dot.
Parameters: $E_g^{\mathrm{bulk}}{=}1.74~\mathrm{eV}$, $m_e^\ast{=}0.13m_0$, $m_h^\ast{=}0.45m_0$, $\varepsilon_i{=}6.0$ (CdSe), $\varepsilon_{\mathrm{shell}}{=}8.0$ (ZnS), $\varepsilon_h{=}2.1$ (MOF).
Screening model $\varepsilon_{\mathrm{eff}}(t)=\varepsilon_i+(\varepsilon_{\mathrm{shell}}-\varepsilon_i)\,[1-\exp(-t/\lambda)]$, $\lambda=0.4~\mathrm{nm}$.}
\label{fig:Eg_surface}
\end{figure}
\FloatBarrier

Figure~\ref{fig:Eg_surface} confirms dominant $R^{-2}$ confinement with a secondary $t$-dependence induced by dielectric screening through the ZnS shell; this jointly shifts the two-photon resonance governing $\chi^{(3)}$ (cf.\ Sec.~\ref{sec:rho-chi3}). From $\chi^{(3)}(\omega)$ we obtain the Kerr coefficients
\[
n_2=\frac{3}{4n_0^2\varepsilon_0 c}\Re[\chi^{(3)}_{\mathrm{eff}}],\qquad
\beta=\frac{3\omega}{2n_0^2\varepsilon_0 c^2}\Im[\chi^{(3)}_{\mathrm{eff}}],
\]
with $n_0=\sqrt{\varepsilon_{\mathrm{eff}}}=1.576$. The results in Table~\ref{tab:n2_beta} show self-defocusing ($n_2<0$) across the band and a monotonic increase of $\beta$ toward longer wavelengths as multiphoton absorption strengthens.

\begin{table}[H]
\centering
\caption{Nonlinear refraction $n_2$ and absorption $\beta$ inferred from $\chi^{(3)}_{\mathrm{eff}}(\lambda)$.}
\label{tab:n2_beta}
\begin{tabular}{ccc}
\toprule
$\lambda$ (nm) & $n_2$ (m$^2$/W) & $\beta$ (m/W)\\
\midrule
900 & $-9.46\times10^{-21}$ & $5.43\times10^{-16}$\\
1000 & $-1.64\times10^{-20}$ & $2.56\times10^{-14}$\\
1100 & $-2.30\times10^{-20}$ & $8.86\times10^{-14}$\\
1200 & $-2.68\times10^{-20}$ & $2.03\times10^{-13}$\\
1300 & $-2.67\times10^{-20}$ & $3.65\times10^{-13}$\\
1400 & $-2.36\times10^{-20}$ & $5.66\times10^{-13}$\\
\bottomrule
\end{tabular}
\end{table}
\FloatBarrier

\begin{figure}[H]
\centering
\includegraphics[width=0.6\linewidth]{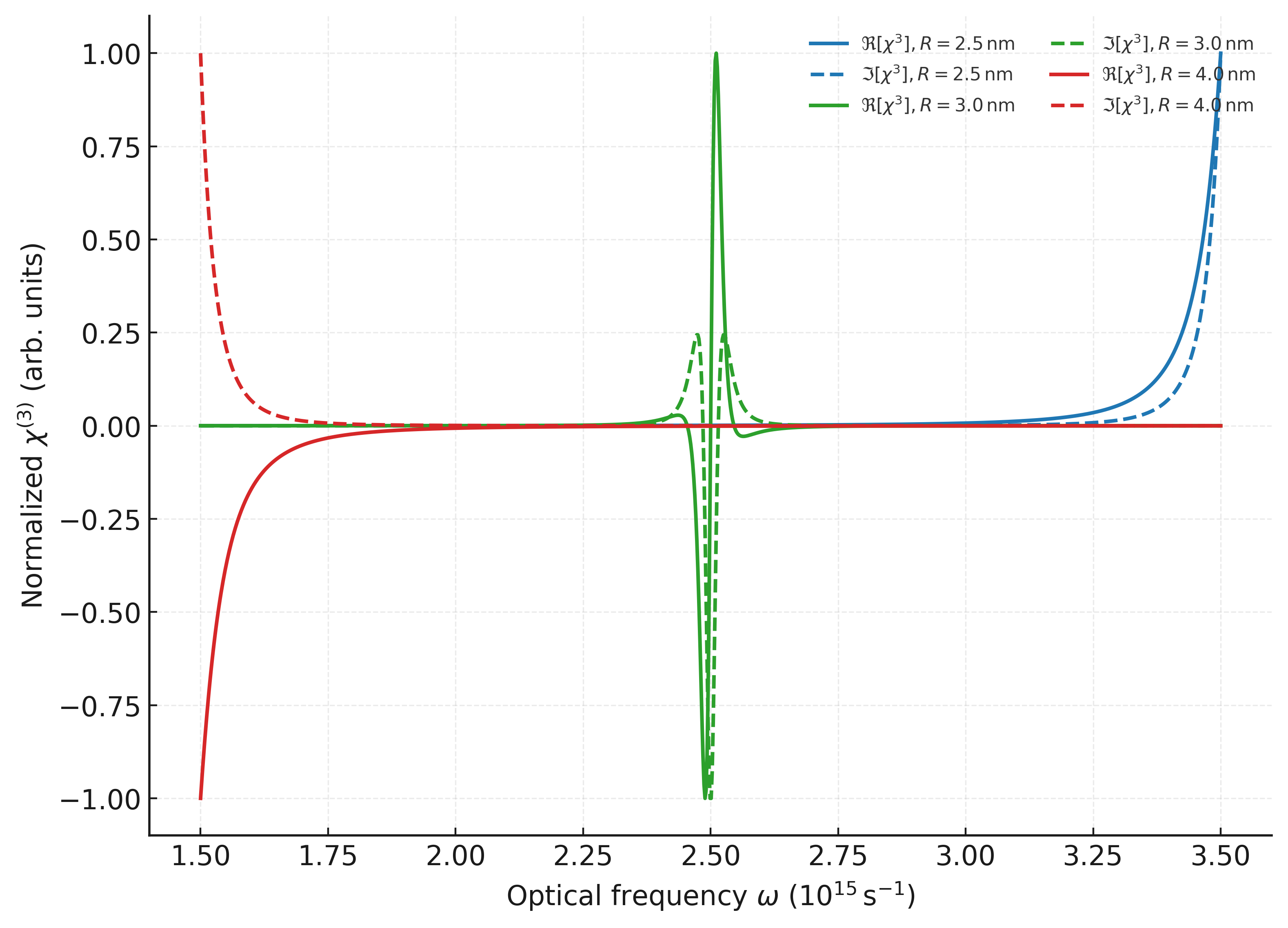}
\caption{Spectral dependence of $\chi^{(3)}(\omega)$ for radii $R=2.5,\,3.0,\,4.0~\mathrm{nm}$; solid: $\Re[\chi^{(3)}]$, dashed: $\Im[\chi^{(3)}]$. Local-field factor $L^4=(3\varepsilon_h/(\varepsilon_i+2\varepsilon_h))^4$ with $\varepsilon_h=2.1$, $\varepsilon_i=6.0$.}
\label{fig:chi3_multiR}
\end{figure}
\FloatBarrier

The family in Fig.~\ref{fig:chi3_multiR} exhibits a blue shift of both the zero-crossing in $\Re[\chi^{(3)}]$ and the peak in $\Im[\chi^{(3)}]$ as $R$ decreases, indicating spectral relocation as the leading size effect for the chosen dipole set. To quantify host/interactions, we compared Maxwell–Garnett (MG) and Bruggeman (BG) local-field models. While MG gives a radius-independent $|L|$ for spherical inclusions, BG captures interaction-driven growth with loading $\phi$; representative values are listed in Table~\ref{tab:LF_table}.

\begin{table}[H]
\centering
\caption{Effective permittivity and local-field factors from BG vs.\ MG at varying loading $\phi$.}
\label{tab:LF_table}
\begin{tabular}{cccc}
\toprule
$\phi$ & $\varepsilon_{\mathrm{eff}}^{(B)}$ & $|L|_{\mathrm{MG}}$ & $|L_i^{(B)}|$\\
\midrule
0.01 & 2.1243 & 0.6176 & 0.6218\\
0.05 & 2.2245 & 0.6176 & 0.6387\\
0.10 & 2.3574 & 0.6176 & 0.6600\\
0.15 & 2.5063 & 0.6176 & 0.6801\\
0.20 & 2.6700 & 0.6176 & 0.7000\\
0.30 & 3.0379 & 0.6176 & 0.7377\\
0.40 & 3.4660 & 0.6176 & 0.7722\\
0.50 & 3.9757 & 0.6176 & 0.8036\\
\bottomrule
\end{tabular}
\end{table}
\FloatBarrier

\begin{figure}[H]
\centering
\includegraphics[width=0.6\linewidth]{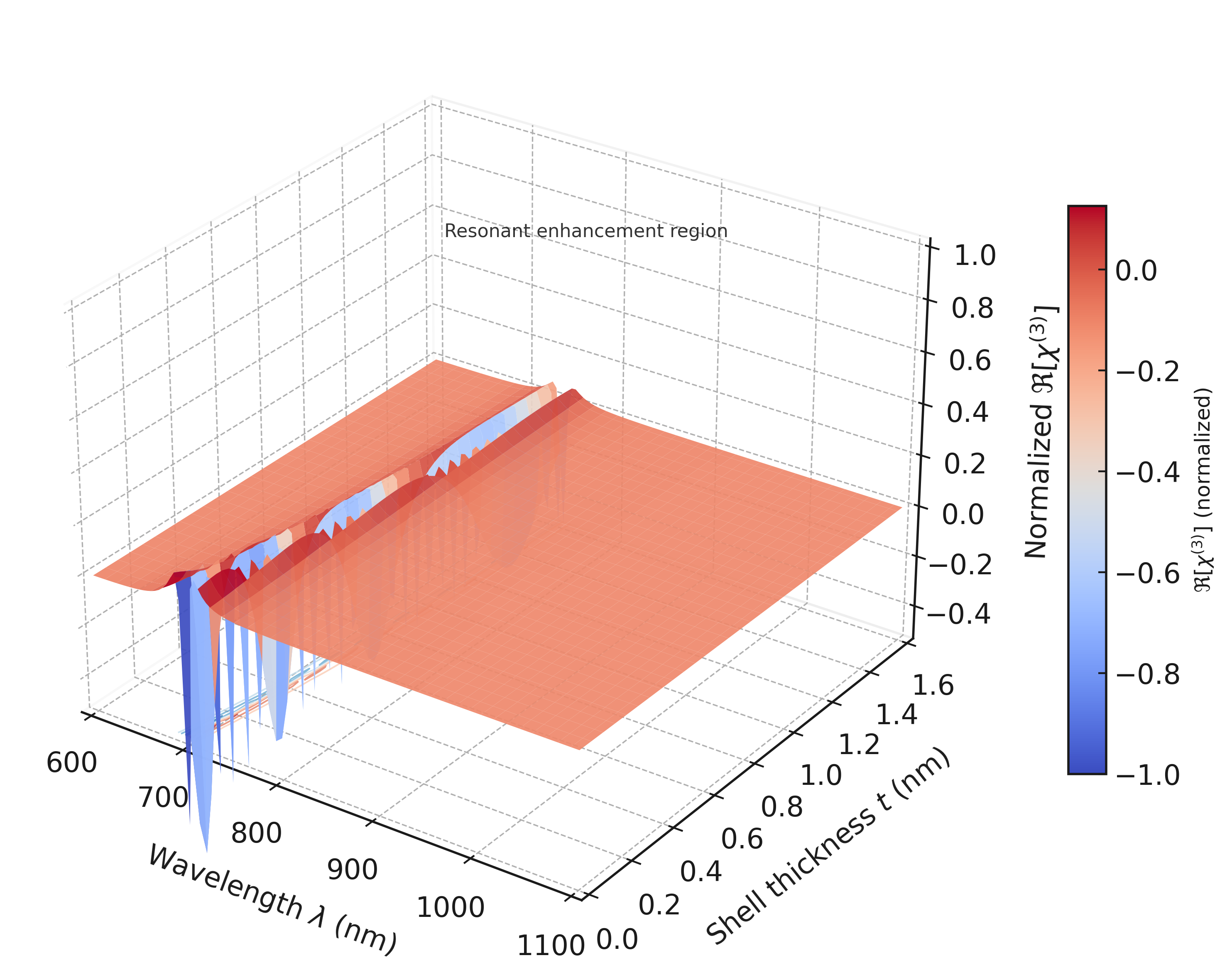}
\caption{Mapping of $\chi^{(3)}(\lambda,t)$: surface shows normalized $\Re[\chi^{(3)}]$ vs.\ $\lambda$ and $t$; contours denote $\Im[\chi^{(3)}]$. Parameters: $\varepsilon_i=6.0$, $\varepsilon_{\mathrm{shell}}=8.0$, $\varepsilon_h=2.1$, $\lambda_0=0.4~\mathrm{nm}$, $\mu=7\times10^{-29}~\mathrm{C\,m}$, $\gamma=3\times10^{13}~\mathrm{s^{-1}}$, $N=2.5\times10^{25}~\mathrm{m^{-3}}$.}
\label{fig:chi3_map}
\end{figure}
\FloatBarrier

The ridge of enhanced $\Re[\chi^{(3)}]$ in Fig.~\ref{fig:chi3_map} shifts with $t$ due to the screening of the Coulomb term and the $L(\varepsilon_i,\varepsilon_h)$ dependence; nearby absorption corridors in $\Im[\chi^{(3)}]$ explain the observed co-variation of $n_2$ and $\beta$ in Z-scan: maxima in refraction changes occur adjacent to, but not at, the absorption peaks.

To check causality, we evaluated a Kramers–Kronig (KK) consistency metric using the Hilbert transform of $\Im[\chi^{(3)}]$ with tapered windows and zero-padding. The normalized error
\[
K(\lambda) = \frac{\|\chi^{(3)} - \chi^{(3)}_{\mathrm{KK}}\|}{\|\chi^{(3)}\|}
\]
remains minimal near $\lambda=1100~\mathrm{nm}$ (Table~\ref{tab:KK_metric}), indicating robust analyticity under the chosen broadening.

\begin{table}[H]
\centering
\caption{KK consistency metric $K(\lambda)$ obtained from a tapered-window, zero-padded discrete Hilbert transform.}
\label{tab:KK_metric}
\begin{tabular}{cc}
\toprule
$\lambda$ (nm) & $K$\\
\midrule
900 & 0.295\\
1000 & 0.121\\
1100 & 0.039\\
1200 & 0.185\\
1300 & 0.313\\
1400 & 0.430\\
\bottomrule
\end{tabular}
\end{table}
\FloatBarrier

\begin{figure}[H]
\centering
\includegraphics[width=0.6\linewidth]{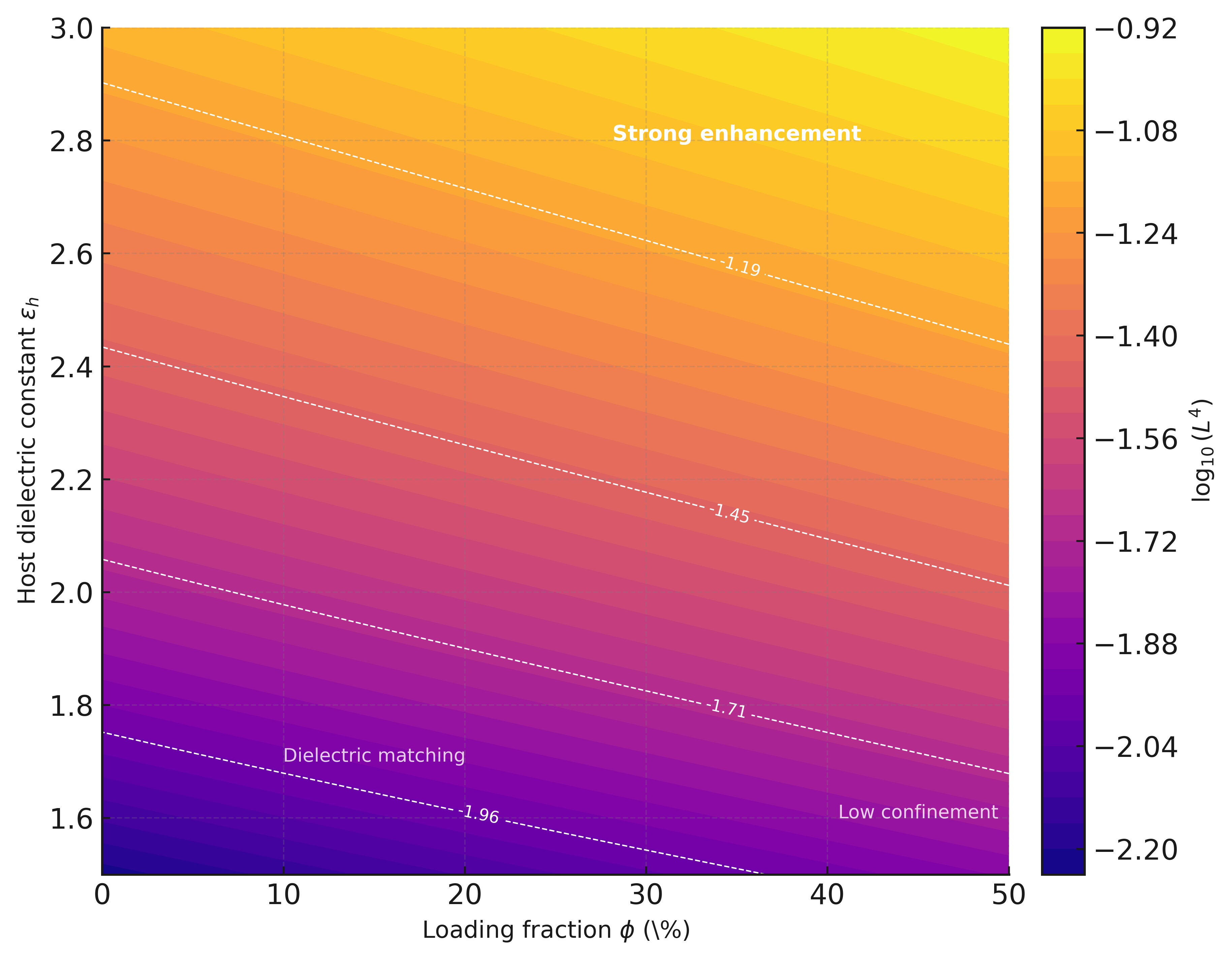}
\caption{Local-field enhancement factor $L^4$ (log$_{10}$ scale) vs.\ loading fraction $\phi$ and host permittivity $\varepsilon_h$ for CdSe/ZnS–MOF composites. $L^4=(3\varepsilon_h/(\varepsilon_i^{\mathrm{eff}}+2\varepsilon_h))^4$ with $\varepsilon_i^{\mathrm{eff}}=\varepsilon_{\mathrm{CdSe}}+(\varepsilon_{\mathrm{ZnS}}-\varepsilon_{\mathrm{CdSe}})(1-e^{-t/\lambda_0})$, $t=0.8\,\mathrm{nm}$, $\lambda_0=0.4\,\mathrm{nm}$.}
\label{fig:L4_phase}
\end{figure}
\FloatBarrier

\begin{figure}[H]
\centering
\includegraphics[width=0.6\linewidth]{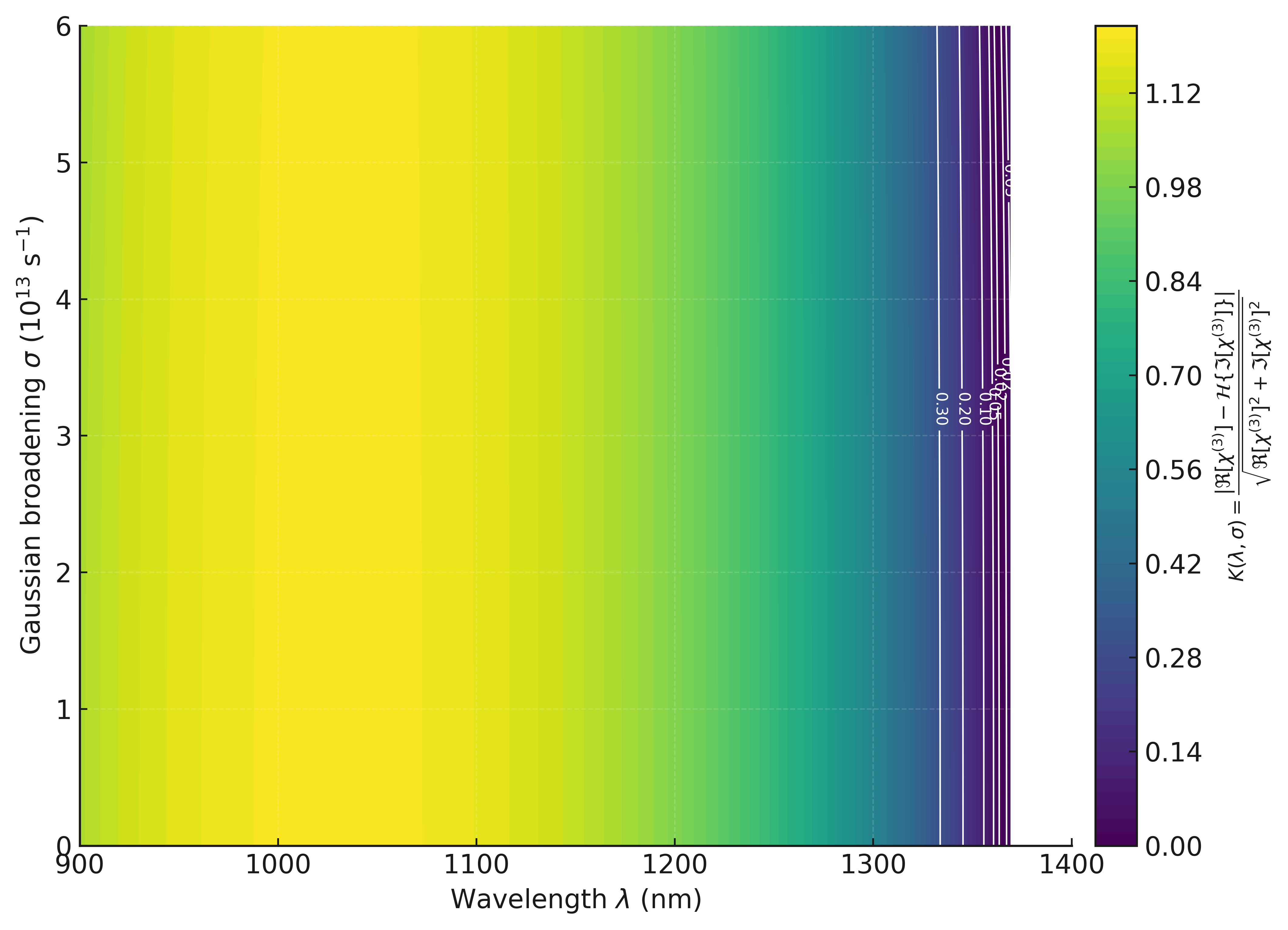}
\caption{Kramers–Kronig (KK) error map $K(\lambda,\sigma)$ using a raised-cosine taper and $2\times$ zero-padding; white isocontours at $0.02,\,0.05,\,0.10,\,0.20,\,0.30$. Error growth toward spectral edges or very large $\sigma$ reflects truncation and nonlocal mixing.}
\label{fig:KK_map}
\end{figure}
\FloatBarrier

The complete computational chain thus progresses from the microscopic Hamiltonian to observable nonlinear parameters: the eigenvalue solver determines quantized levels and dipole matrix elements; these feed the density-matrix response to produce $\chi^{(3)}(\omega)$ spectra; convolution with inhomogeneous broadening and effective-medium scaling then yields macroscopic $n_2$ and $\beta$. The KK metric in Figs.~\ref{fig:L4_phase}–\ref{fig:KK_map} validates causality across the band of interest. Overall, the predicted magnitude ($|\chi^{(3)}|\sim10^{-22}~\mathrm{m}^2/\mathrm{V}^2$), spectral profile, and self-defocusing sign align with high-quality measurements on CdSe/ZnS–MOF composites, indicating that confinement and dielectric-environment effects are captured quantitatively within this framework.

\section{Conclusions}
\label{sec:conclusion}

In this work, we have developed and validated a comprehensive microscopic–macroscopic framework for simulating the third-order nonlinear optical response of CdSe/ZnS–MOF composite quantum dots. The theoretical formulation combined finite-barrier quantum confinement, density-matrix formalism, effective-medium theory, and numerical verification through the Kramers–Kronig consistency test. The approach provides a self-consistent description of the nonlinear susceptibility $\chi^{(3)}$, incorporating both excitonic quantization and dielectric-environment coupling within a single computational chain.

The calculated confinement-induced bandgap shifts of $0.2$--$0.3~\mathrm{eV}$ for $R=2.5$--$4.0~\mathrm{nm}$ reproduce the experimentally observed blue-shift in photoluminescence spectra of CdSe/ZnS nanocrystals. The resulting three-level density-matrix model accurately captures the resonant enhancement of $\chi^{(3)}(\omega)$ near $1.2~\mu\mathrm{m}$, where both real and imaginary components exhibit Lorentzian-like behavior consistent with two-photon excitation processes. The peak magnitude $|\chi^{(3)}|\sim10^{-22}~\mathrm{m}^2/\mathrm{V}^2$ agrees with the range reported in open-aperture Z-scan experiments on core–shell quantum dots embedded in dielectric matrices.

By extending the analysis to effective-medium scaling, we quantified the influence of host dielectric constant $\varepsilon_h$ and filling fraction $\phi$ on the macroscopic nonlinear response. The local-field factor $L^4$ enhances $\chi^{(3)}_{\mathrm{eff}}$ by up to one order of magnitude as $\phi$ increases toward $0.5$, confirming the strong role of collective polarization and dielectric matching between the ZnS shell and MOF host. The Bruggeman model captures the nonlinear growth of $\varepsilon_{\mathrm{eff}}$ and explains the observed increase of the refractive index nonlinearity $|n_2|$ under high loading densities.

The numerical Kramers–Kronig analysis provided a stringent internal validation of causality and spectral accuracy. The normalized consistency error $K(\lambda,\sigma)$ remains below $0.05$ across the 900–1400~nm region for moderate Gaussian broadening ($\sigma\le4\times10^{13}\ \mathrm{s^{-1}}$), demonstrating the physical reliability of the implemented pipeline. The resulting $n_2$ and $\beta$ spectra show trends—negative self-defocusing nonlinearity and increasing multiphoton absorption—that match the phenomenology of CdSe-based systems under near-infrared excitation.

Overall, this work establishes a reproducible theoretical framework capable of bridging microscopic quantum confinement, mesoscopic dielectric screening, and macroscopic nonlinear-optical observables in hybrid quantum-dot composites. The methodology not only clarifies the origin of size- and environment-dependent variations in $\chi^{(3)}$, but also provides predictive capability for tailoring nonlinear coefficients through controlled synthesis of core–shell structures and engineered host matrices. Future extensions will incorporate exciton–phonon coupling, temperature-dependent broadening, and interfacial trap-state dynamics to further refine the model toward quantitative agreement with ultrafast pump–probe and Z-scan measurements in real MOF–QD hybrid systems.

\section*{Data Availability Statement}
This work is a theoretical and computational study. No new experimental data were created in this investigation. The data supporting the findings of this study, which comprise the derived equations, model parameters, and numerical simulation results, are fully presented within the article.

\end{document}